\documentclass[aps, prd, twocolumn, superscriptaddress, nofootinbib, floatfix, noshowpacs]{revtex4-2}
\pdfoutput=1
\newif\ifcomment
\usepackage{amsmath,amssymb}
\usepackage{color,url}
\usepackage{listings}
\usepackage{slashed}
\usepackage[pdftex]{graphicx}
\usepackage{epstopdf}
\usepackage{epsfig}
\usepackage{grffile}
\usepackage{relsize}
\usepackage{float}
\graphicspath{{./img/}}
\usepackage{soul}

\usepackage{CJKutf8}

\usepackage[mathscr,scaled=1.15]{urwchancal}
\DeclareFontFamily{OT1}{pzc}{}
\DeclareFontShape{OT1}{pzc}{m}{it}%
{<-> s * [1.15] pzcmi7t}{}
\DeclareMathAlphabet{\mathpzc}{OT1}{pzc}{m}{it}

\definecolor{purple}{rgb}{0.5,0,0.5}
\definecolor{blue}{rgb}{0.0,0,0.9}
\definecolor{prdblue}{rgb}{0.133,0.118,0.498}
\usepackage[colorlinks=true, pdfstartview=FitV, linkcolor=prdblue, citecolor= prdblue, urlcolor=prdblue]{hyperref}

\RequirePackage{comment}

\newcommand{\beq}{\begin{equation}}
\newcommand{\eeq}{\end{equation}}
\newcommand{\ba}{\begin{array}}
\newcommand{\ea}{\end{array}}
\newcommand{\bea}{\begin{align}}
\newcommand{\eea}{\end{align}}
\newcommand{\bi}{\begin{itemize}}
\newcommand{\ei}{\end{itemize}}
\newcommand{\ben}{\begin{enumerate}}
\newcommand{\een}{\end{enumerate}}
\newcommand{\bc}{\begin{center}}
\newcommand{\ec}{\end{center}}
\newcommand{\bl}{\begin{flushleft}}
\newcommand{\el}{\end{flushleft}}
\newcommand{\br}{\begin{flushright}}
\newcommand{\er}{\end{flushright}}

\begin{document}
\begin{CJK*}{UTF8}{gbsn}

\title{$\,$\\[-6ex]\hspace*{\fill}{\normalsize{\sf\emph{Preprint no}.\
NJU-INP 106-25}}\\[1ex]
Symmetry Constraints on Pion Valence Structure}

\author{Xiaobin Wang (王晓斌)%
$^ {\href{https://orcid.org/0000-0002-2786-5296}{\textcolor[rgb]{0.00,1.00,0.00}{\sf ID}}\,}$}
\affiliation{School of Physics, \href{https://ror.org/01y1kjr75}{Nankai University}, Tianjin 300071, China}

\author{Lei Chang (常雷)%
$^ {\href{https://orcid.org/0000-0002-4339-2943}{\textcolor[rgb]{0.00,1.00,0.00}{\sf ID}}\,}$}
\email{leichang@nankai.edu.cn}
\affiliation{School of Physics, \href{https://ror.org/01y1kjr75}{Nankai University}, Tianjin 300071, China}

\author{\\Minghui Ding (丁明慧)%
    $^{\href{https://orcid.org/0000-0002-3690-1690}{\textcolor[rgb]{0.00,1.00,0.00}{\sf ID}}\,}$}
\email[]{mhding@nju.edu.cn}
\affiliation{School of Physics, \href{https://ror.org/01rxvg760}{Nanjing University}, Nanjing, Jiangsu 210093, China}
\affiliation{Institute for Nonperturbative Physics, \href{https://ror.org/01rxvg760}{Nanjing University}, Nanjing, Jiangsu 210093, China}

\author{Kh\'epani Raya%
    $^{\href{https://orcid.org/0000-0001-8225-5821}{\textcolor[rgb]{0.00,1.00,0.00}{\sf ID}}\,}$}
\affiliation{Department of Integrated Sciences and Center for Advanced Studies in Physics, Mathematics and Computation, \href{https://ror.org/03a1kt624}{University of Huelva}, E-21071 Huelva, Spain.}

\author{Craig D. Roberts%
       $^{\href{https://orcid.org/0000-0002-2937-1361}{\textcolor[rgb]{0.00,1.00,0.00}{\sf ID}}\,}$}
\email[]{cdroberts@nju.edu.cn}
\affiliation{School of Physics, \href{https://ror.org/01rxvg760}{Nanjing University}, Nanjing, Jiangsu 210093, China}
\affiliation{Institute for Nonperturbative Physics, \href{https://ror.org/01rxvg760}{Nanjing University}, Nanjing, Jiangsu 210093, China}
%

\date{
2026 June 24} 

\begin{abstract}
The profile of the pion valence quark distribution function (DF) remains controversial.
Working from the concepts of QCD effective charges and generalised parton distributions, we show that since the pion elastic electromagnetic form factor is well approximated by a monopole, then, at large light-front momentum fraction, the pion valence quark DF is a convex function described by a large-$x$ power law that is practically consistent with expectations based on quantum chromodynamics.\\[1ex]
\end{abstract}

\maketitle

\end{CJK*}

\noindent\emph{1.\ Introduction}\,---\,%
Pions are the messengers of Nature's strong interaction, being responsible for the long-range binding that stabilises all nuclei.
There are three pions: $\pi^\pm, \pi^0$.
They form an isospin triplet.
The $\pi^\pm$ are a both particle antiparticle pairs; hence, they are mass degenerate.
The $\pi^0$ is its own antiparticle and is almost mass degenerate with $\pi^\pm$.
The small splitting, $[m_{\pi^\pm}-m_{\pi^0}]/m_{\pi^\pm} = 3.3$\% \cite{ParticleDataGroup:2024cfk}, owes to isospin-symmetry breaking in the strong interaction, which is small, and electromagnetic contributions.

In effective theories of nuclear interactions, pions are typically considered to be point particles \cite{Lynn:2019rdt, Launey:2021sua, Ananthanarayan:2023gzw}.
However, they are not.
Pions exhibit a special dichotomy.
Supposing that quantum chromodynamics (QCD) is the theory underlying strong interactions \cite{Fritzsch:1973pi},
then pions are both  \cite{Lane:1974he, Maris:1997tm, Llanes-Estrada:2001bgw, Bicudo:2001jq, Qin:2014vya}: (\emph{a}) the Goldstone bosons that emerge as a consequence of dynamical chiral symmetry breaking in QCD and (\emph{b}) QCD bound states, seeded by a light valence quark ($u$ -- up; $d$ -- down) and a light valence antiquark.

Pion structure was first explored in experiments that determined the $\pi^\pm$ charge radius \cite{Dally:1982zk, Amendolia:1984nz, Amendolia:1986wj}: its radius is roughly 80\% of that of the proton \cite{ParticleDataGroup:2024cfk}, so the pion is far from pointlike.
Interior structural information has been obtained using $\pi +W$ Drell-Yan reactions \cite[E615]{Conway:1989fs, Wijesooriya:2005ir} and deep-inelastic scattering (DIS) reactions $e p \to e^\prime \pi^+ n $, \emph{i.e}., DIS on a proton target with a tagged neutron in the final state \cite{Derrick:1996ax, Adloff:1998yg}.

Only the E615 data provide information on pion valence quark structure and analyses of that data continue to be controversial \cite{Cui:2021mom, dePaula:2022pcb, Ahmady:2022dfv, Chang:2022pcb, Pasquini:2023aaf, Alberg:2024svo, Choi:2024ptc, Lan:2024ais, Lin:2025hka}.  The data are intended to inform our understanding of the valence quark distribution function (DF), ${\mathpzc l}_\pi(x;\zeta)$, which is a nonnegative density that describes the probability that a given valence quark carries a light-front fraction of the pion's momentum in the range $[x+dx,x]$ when perceived by a probe with resolving energy scale $\zeta$.
Herein, we deliver a collection of model-independent and parameter-free predictions that constrain the behaviour of the pion valence quark DF.

\smallskip

\noindent\emph{2.\ Hadron scale and DF symmetries}\,---\,%
We begin with the concept of a QCD effective charge \cite{Grunberg:1980ja, Grunberg:1982fw}.
In this scheme, a QCD running coupling is defined using the expansion of a chosen observable restricted to first order in the perturbative coupling, $\alpha_s$; see the discussion in Ref.\,\cite[Sec.\,4.3]{Deur:2023dzc}.
Consequently, effective charges are typically process dependent, like $\alpha_{g_1}(k^2)$, defined via the Bjorken sum rule \cite{Deur:2022msf}.
Significantly, any coupling obtained in this way is analogous to the Gell-Mann--Low coupling in quantum electrodynamics \cite{GellMann:1954fq}.
Moreover, it is: consistent with the QCD renormalisation group; renormalisation scheme independent; everywhere analytic and finite; and supplies an infrared completion of any standard (perturbatively defined) running coupling.

Herein, we elect to work with an effective charge, $\alpha_{1\ell}(k^2)$, that is a function which, when used to integrate the leading-order perturbative DGLAP equations \cite{Dokshitzer:1977sg, Gribov:1971zn, Lipatov:1974qm, Altarelli:1977zs}, defines an evolution scheme for all parton DFs -- both unpolarised and polarised, and for any hadron -- that is all-orders exact.
This is the all-orders (AO) approach \cite{Yin:2023dbw}.
The definition is broader than usual because it refers to an entire class of observables, not just a single measurable quantity.
It is important to stress that the pointwise form $\alpha_{1\ell}(k^2)$ is largely irrelevant: a large array of model-independent results can be proved without any reference to the $k^2$-profile of the effective charge \cite{Yin:2023dbw}.

Suppose now that one has a $\pi^+$ valence quark DF at some scale $\zeta \gg m_p$, where $m_p$ is the proton mass, \emph{i.e}.,
${\mathpzc u}_\pi(x,\zeta)$, where $x\in[0,1]$.
Owing to ${\cal G}$-parity symmetry \cite{Lee:1956sw}, which is a good approximation in Nature, then \cite{Hecht:2000xa}:
\begin{equation}
{\mathpzc u}_{\pi^+}(x,\zeta) = \bar {\mathpzc d}_{\pi^+}(x;\zeta) \,,
\label{piGparity}
\end{equation}
with similar statements for $\pi^{-,0}$.  

Given the existence of $\alpha_{1\ell}(k^2)$, which is available at all $k^2$, by definition, a next step is possible.
Namely, one can use $\alpha_{1\ell}(k^2)$-extended DGLAP evolution, \emph{viz}.\ the AO method, to evolve ${\mathpzc u}_{\pi^+}(x,\zeta)$ to a scale $\zeta_{\cal H}\lesssim m_p$, whereat all properties of the pion -- including its light-front momentum -- are carried by its valence degrees of freedom.
This is the hadron scale, $\zeta_{\cal H}$.
The numerical value of $\zeta_{\cal H}$ is immaterial.  One need only know that it exists; a fact guaranteed by the theory of effective charges.

At $\zeta_{\cal H}$, the valence-quark and -antiquark are quasiparticles, into which all glue and sea partons have been sublimated \cite{Yao:2025fnb}.
This is just a statement of dressing and is expressed analogously in any QCD Schwinger function; consider, \textit{e.g}., a diagrammatic expansion of the quark gap equation.
Exploiting Eq.\,\eqref{piGparity}, it follows that
\begin{equation}
{\mathpzc u}_{\pi^+}(x,\zeta_{\cal H}) = \bar {\mathpzc d}_{\pi^+}(x)
= {\mathpzc u}_{\pi^+}(1-x,\zeta_{\cal H})\,,
\label{DFsym}
\end{equation}
because symmetry requires that the light-front momentum be shared equally between the valence constituents.  Thus, at the hadron scale, there is only one nontrivial DF, which is that of the valence $u$ quark in the $\pi^+$.

As explained elsewhere \cite{Yin:2023dbw}, these statements are true by definition.  The issue is whether anything useful can therefrom be said about hadron observables.  Today, one can answer in the affirmative because this approach has proved efficacious in many applications, providing, \emph{e.g}.,
unified predictions for all -- including glue and sea -- pion and kaon DFs \cite{Cui:2020tdf};
all proton, $\Lambda$, $\Sigma^0$ DFs \cite{Chang:2022jri, Lu:2022cjx, Cheng:2023kmt, Yu:2024qsd, Yu:2025fer};
insights from experiment into such DFs \cite{Xu:2023bwv, Xu:2024nzp};
useful information on quark and gluon angular momentum contributions to the proton spin \cite{Yu:2024ovn};
predictions for $\pi$ and $K$ fragmentation functions \cite{Xing:2025eip};
and a tenable species separation of nucleon gravitational form factors \cite{Yao:2024ixu, Binosi:2025kpz}.

Using the overlap representation of generalised parton distributions (GPDs) \cite{Diehl:2000xz}, one has the following expression for the pion valence quark GPD:
\begin{align}
H^{{\mathpzc u}_\pi}& (x,Q^2;\zeta_{\cal H}) \nonumber\\
&  = \int d^2 \,b_\perp \, {\rm e}^{i (1-x) b_\perp \cdot Q}
| \psi^{{\mathpzc u}_\pi}(x,b_\perp;\zeta_{\cal H})|^2,
\label{EqGPD}
\end{align}
where $\psi^{{\mathpzc u}_\pi}(x,b_\perp;\zeta_{\cal H})$ is the pion valence-quark light-front wave function in impact parameter space \cite{Burkardt:2002hr}: $b_\perp$ is Fourier conjugate to $k_\perp$, the hadron momentum in the light-front transverse plane.
(Herein, we need only consider zero skewness.)
Canonical normalisation requires
\begin{equation}
\int_{0}^1dx\int d^2 b_\perp | \psi^{{\mathpzc u}_\pi}(x,b_\perp;\zeta_{\cal H})|^2 = 1\,.
\end{equation}
Furthermore and importantly \cite{Diehl:2003ny}:
\begin{equation}
{\mathpzc u}_{\pi}(x,\zeta_{\cal H}) = H^{{\mathpzc u}_\pi} (x,Q^2=0;\zeta_{\cal H})\,,
\label{hpiHpi}
\end{equation}
\emph{i.e}., the pion valence quark DF is the $Q^2=0$ (forward) limit of the pion valence quark GPD.
Such DFs are nonnegative functions.

In the AO evolution setting, there is no other component of $\psi^{{\mathpzc u}_\pi}(x,{b}_\perp;\zeta_{\cal H})$ at $\zeta_{\cal H}$.
Thus, symmetry entails
\begin{equation}
\psi^{{\mathpzc u}_\pi}(x,b_\perp;\zeta_{\cal H})
= \psi^{{\mathpzc u}_\pi}(1-x,b_\perp;\zeta_{\cal H})\,;
\end{equation}
hence, using Eq.\,\eqref{EqGPD},
\begin{equation}
H^{{\mathpzc u}_\pi} (x,Q^2;\zeta_{\cal H})
= H^{{\mathpzc u}_\pi} (1-x,Q^2 [1-x]^2/x^2;\zeta_{\cal H}) \,.
\label{GPDsym}
\end{equation}

Thus far, only ${\cal G}$-parity symmetry has been used.

\smallskip

\noindent\textit{3.\ Constraints via pion electromagnetic form factor}\,---\,
The pion form factor, $F_\pi(Q^2)$, has been measured \cite{Dally:1982zk, Amendolia:1984nz, Amendolia:1986wj, Horn:2007ug, Huber:2008id}.
Existing data, available on $0<Q^2/{\rm GeV}^2\leq 2.45$, are well approximated by the following monopole function \cite{Yao:2024drm}:
\begin{equation}
F_\pi^{M}(Q^2)=1/(1+Q^2/m_M^2)\,,\quad m_M \approx 0.73\,{\rm GeV}.
\label{MonoAssume}
\end{equation}

On the other hand, perturbative QCD (pQCD) predicts \cite{Lepage:1979zb, Efremov:1979qk, Lepage:1980fj} that on some domain $Q^2 \gg m_p^2$, $F_\pi(Q^2)$ falls faster than a monopole, with the additional logarithmic suppression expressing the impact of anomalous dimensions in QCD.
Continuum analyses predict that the difference between $Q^2 F_\pi(Q^2)$ and $Q^2F_\pi^{M}(Q^2)$ becomes noticeable on $Q^2 \gtrsim 5\,$GeV$^2$ \cite{Yao:2024drm}.
However, the difference is only visible because $Q^2$ becomes large.

If one does not multiply by $Q^2$, then, quantitatively, $F_\pi^{M}(Q^2)$ is a fair representation of $F_\pi(Q^2)$ on the entire domain of spacelike momenta because the difference between these functions vanishes as $1/Q^2$.  We will thus proceed by assuming that $F_\pi^{M}(Q^2)$ is a practicable approximation to $F_\pi(Q^2)$, and use the following fact \cite{Chang:2020kjj}:
\begin{equation}
F_\pi^{M}(Q^2)
= (1/2) B(1, 1/2 + Q^2/[2m_M^2])\,,
\label{FpiApp}
\end{equation}
where the Euler $\beta$-function is defined thus:
\begin{equation}
B(1+a,1+b) = \int_0^1 dy\, (1-y)^a y^b .
\label{EqEB}
\end{equation}

In QCD, Eq.\,\eqref{FpiApp} and its corollaries (to be explored below) receive small modifications at large $Q^2$ and, consequently, on the endpoint domains $x\simeq 0,1$.

The pion form factor can also be expressed in terms of the pion hadron-scale valence quark GPD \cite{Diehl:2003ny}:
\begin{equation}
F_\pi(Q^2) =
\int_0^1 dx H^{{\mathpzc u}_\pi} (x,Q^2;\zeta_{\cal H})\,,
\label{FpiGPD}
\end{equation}
where we have used $G$-parity symmetry .

The similarity between Eqs.\,\eqref{FpiApp}, \eqref{EqEB} and \eqref{FpiGPD} means that, by exploiting properties of integrable, non-negative and continuously differentiable ($C^1$) functions on $x\in [0,1]$, it is readily shown that there is a class of $Q^2$-independent $C^1$ functions, $\{w(x)$\}, satisfying
\begin{equation}
\label{wcons}
w(0)=0\,, \quad w(1)=1 \,, \quad w^\prime(x)\geq0\,,
\end{equation}
in terms of which one may write \cite{Wang:2024fjt}
\begin{subequations}
\label{LFHMGPD}
\begin{align}
H^{{\mathpzc u}_\pi} (x,Q^2;\zeta_{\cal H})
& = \tilde{\mathpzc u}_{\pi}(x;\zeta_{\cal H}) [w(x)]^{Q^2/2m_M^2} \,,
\label{dewA}\\
\tilde{\mathpzc u}_{\pi}(x;\zeta_{\cal H}) & =
\tfrac{1}{2} [w(x)]^{-1/2} w^\prime (x)
=: {\mathpzc a}^\prime (x)
\,. \label{dew}
\end{align}
\end{subequations}

Now applying the forward-limit GPD constraint in Eq.\,\eqref{hpiHpi}, one selects a single member of the class, \emph{i.e}., the function $w(x)$ for which
{\allowdisplaybreaks
\begin{align}
{\mathpzc u}_{\pi}(x;\zeta_{\cal H}) &
 = \tilde{\mathpzc u}_{\pi}(x;\zeta_{\cal H}) =  {\mathpzc a}^\prime (x)
 %
\label{DFw}
\end{align}
and thereby arrives at a minimal expression for
$H^{{\mathpzc u}_\pi} (x,Q^2;\zeta_{\cal H})$
that complies with Eqs.\,\eqref{hpiHpi}, \eqref{FpiGPD}.
}

We stress that Eq.\,\eqref{DFw} is valid for any bound state whose electromagnetic form factor is well approximated by a monopole, irrespective of the $m_M$-value.

\smallskip

\noindent\emph{4.\ Symmetry and the pion $\zeta_{\cal H}$ valence quark GPD}\,---\,%
The preceding analysis has established that:
\begin{equation}
H^{{\mathpzc u}_\pi} (x,Q^2;\zeta_{\cal H})
= {\mathpzc u}_{\pi}(x;\zeta_{\cal H}) [{\mathpzc a}(x)]^{Q^2/m_M^2} \,,
\label{Hfinal}
\end{equation}
where, using Eq.\,\eqref{DFw}, the function ${\mathpzc a}(x)$ is completely determined by the hadron-scale valence quark DF or, alternatively, with this function in hand, one delivers a prediction for that DF.
The rate decreases as $m_M$  increases; so, more pointlike bound states are associated with $Q^2$-flatter GPDs.

\begin{figure}[t]
\includegraphics[width=0.93\columnwidth]{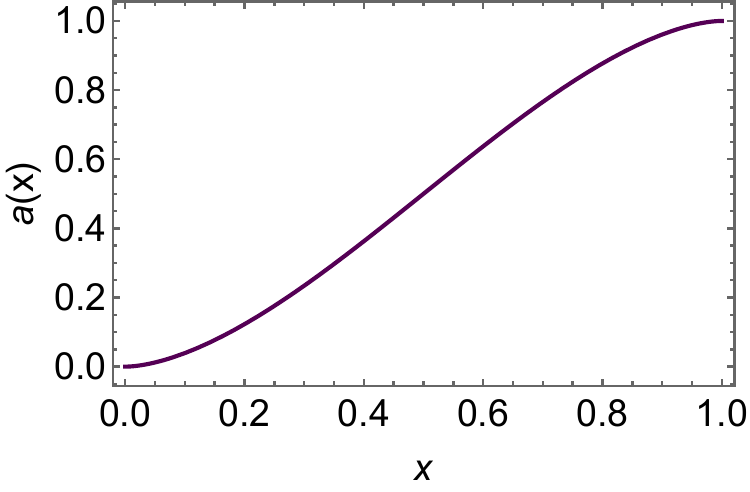}
\caption{\label{Fig1}
Solution of Eq.\,\eqref{EqMaster}, which satisfies the requirements imposed by Eq.\,\eqref{wcons}.
}
\end{figure}

Defining ${\mathpzc m}(x) = (1/m_M^2)\ln [1/{\mathpzc a}(x)]$, it is useful to work with the following form of Eq.\,\eqref{Hfinal} \cite{Raya:2024glv}:
\begin{equation}
H^{{\mathpzc u}_\pi} (x,Q^2;\zeta_{\cal H})
= {\mathpzc u}_{\pi}(x;\zeta_{\cal H})
\exp[-Q^2 {\mathpzc m}(x)]\,.
\end{equation}
For instance, one may now readily obtain
\begin{align}
|\psi(x,b_{\perp};\zeta_{\cal H})|^2
& =\frac{{\mathpzc u}_\pi(x;\zeta_{\cal H})}{4\pi {\mathpzc m}(x)} \nonumber \\
& \quad \times  (1-x)^2
\exp\left[-\frac{(1-x)^2 b_\perp^2}{4 {\mathpzc m}(x)}\right]\,.
\label{HSGPDb}
\end{align}
{\allowdisplaybreaks
From this expression, the distribution amplitude (DA) associated with the pion's two dressed valence parton component is directly obtain via
\begin{subequations}
\label{pionDA}
\begin{align}
\varphi_\pi(x;\zeta_{\cal H}) & =
2\pi |\psi(x,b_{\perp}=0;\zeta_{\cal H})|/m_M \\
& = (1-x) (\pi {\mathpzc u}_{\pi}(x;\zeta_{\cal H}) /\ln[1/{\mathpzc a}(x)])^{\tfrac{1}{2}}\,.
\end{align}
\end{subequations}}

Now, using Eq.\,\eqref{wcons}, the DF symmetry property, Eq.\,\eqref{DFsym}, entails
\begin{equation}
{\mathpzc a}(x)+{\mathpzc a}(1-x) = 1\,.
\label{asymm}
\end{equation}
Subsequently using the GPD symmetry property, Eq.\,\eqref{GPDsym}, one arrives at the following ``master equation'':
\begin{equation}
1 - {\mathpzc a}(x) = [{\mathpzc a}(x)]^{x^2/(1-x)^2}\,.
\label{EqMaster}
\end{equation}

\begin{figure}[t]
\includegraphics[width=0.93\columnwidth]{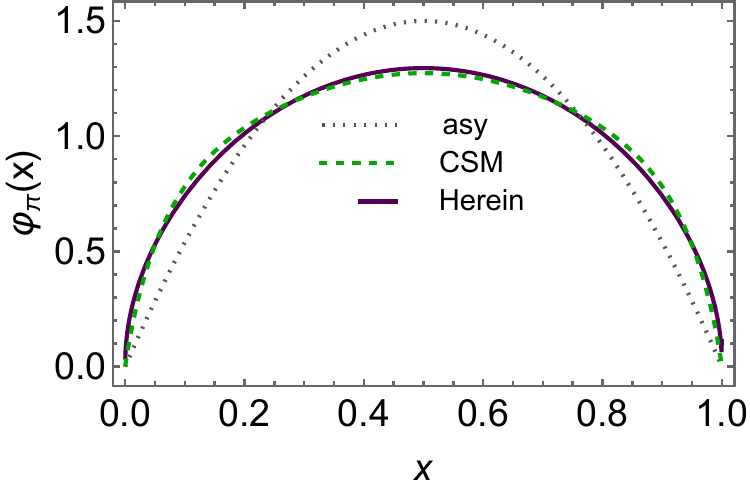}
\caption{\label{Fig2}
Pion valence quark DA, Eq.\,\eqref{pionDA}.
Legend:
solid purple curve -- result obtained herein using ${\mathpzc a}(x)$ from Fig.\,\ref{Fig1} and Eq.\,\eqref{DFw};
dashed green curve -- CSM prediction \cite{Chang:2013pq, Cui:2020tdf};
dotted grey curve -- asymptotic twist-two pion DA \cite{Lepage:1979zb, Efremov:1979qk, Lepage:1980fj}.
}
\end{figure}

It is worth reiterating that this identity is valid for any bound state whose electromagnetic form factor is well approximated by a monopole, irrespective of the $m_M$-value.  No other dynamical information is required to fix the form of ${\mathpzc a}(x)$ and, hence, the pion hadron scale valence quark GPD.

\smallskip

\noindent\emph{5.\ Distribution amplitudes and functions}\,---\,%
The solution of Eq.\,\eqref{EqMaster} is drawn in Fig.\,\ref{Fig1}.
For practical applications, it is well approximated by the following polynomial:
\begin{equation}
{\mathpzc a}(x) \approx 0.195 x + 2.415 x^2 - 1.610 x^3\,.
\end{equation}
Nevertheless, herein, we employ the numerically obtained result; and, for future use, note that the true large-$x$ behaviour is as follows:
\begin{align}
{\mathpzc a}(x) & \stackrel{x\simeq 1}{=}  1  - 1.581 (1-x)^2 \ln 1/(1-x) \nonumber \\
& \quad - 5.458 (1-x)^4 [\ln 1/(1-x)]^2 + \ldots\ ,
\label{aendpt}
\end{align}
with the profile on $x\simeq 0$ obtained via Eq.\,\eqref{asymm}.

Using the solution of Eq.\,\eqref{EqMaster}, one obtains the pion valence quark DA drawn in Fig.\,\ref{Fig2}.
It has the following endpoint behaviour, obtained with Eqs.\,\eqref{DFw}, \eqref{pionDA}, \eqref{aendpt}:
\begin{equation}
\varphi_\pi(x) \stackrel{x\simeq 1}{=} 2.507 \sqrt{1-x}+\ldots\,.
\label{phiendpt}
\end{equation}
The $x\simeq 0$ form is given by $x \leftrightarrow (1-x)$ symmetry; and
$\int_0^1 dx (2x-1)^2 \varphi_\pi(x) =0.246=:\langle \xi^2 \rangle_{\varphi_\pi}$.

The solution is compared in Fig.\,\ref{Fig2} with the continuum Schwinger function method (CSM) prediction \cite{Chang:2013pq, Cui:2020tdf}, obtained using a Bethe-Salpeter kernel for the pion bound state equation which maintains a rigorous link with pQCD and, consequently, ensures that scaling violations are expressed in the pion elastic electromagnetic form factor; see also Ref.\,\cite{Yao:2024drm}. 
The CSM DA prediction is almost indistinguishable from the symmetry-driven result obtained herein, \emph{e.g}., $\langle \xi^2 \rangle_{\varphi_\pi^{\rm CSM}}=0.247$.
Those small differences which do exist are merely an expression of the absence of scaling violations in the monopole approximation to $F_\pi(Q^2)$, Eq.\,\eqref{FpiApp}.
(One may verify this by changing the monopole power from $1\to 1+\epsilon$, with $\epsilon \simeq 0$ serving to mockup a $\ln$-correction; see the Appendix.)
This is also reflected in a different endpoint behaviour:
$\varphi_\pi^{\rm CSM} \approx 20.28 (1-x)$ on $x\simeq 1$, again with the $x\simeq 0$ result given by $x \leftrightarrow (1-x)$ symmetry, and again with endpoint power-law behaviour that is consistent with QCD expectations.
The remaining curve in Fig.\,\ref{Fig2} is the pQCD prediction for the asymptotic twist-two pion DA, valid for $\zeta$ in the neighbourhood $\zeta_{\cal H}/\zeta \simeq 0$:
$\varphi_{\rm asy} = 6 x (1-x)$, $\langle \xi^2 \rangle_{\varphi_{\rm asy}}=0.2$.

Plainly, the symmetry properties discussed in Sec.\,2 entail that, for any meson whose electromagnetic form is well approximated by a monopole, the DA associated with the two dressed valence parton component is a dilated, concave function, much broader and flatter than the asymptotic DA.
This is a model-independent result.

\begin{figure}[t]
\includegraphics[width=0.93\columnwidth]{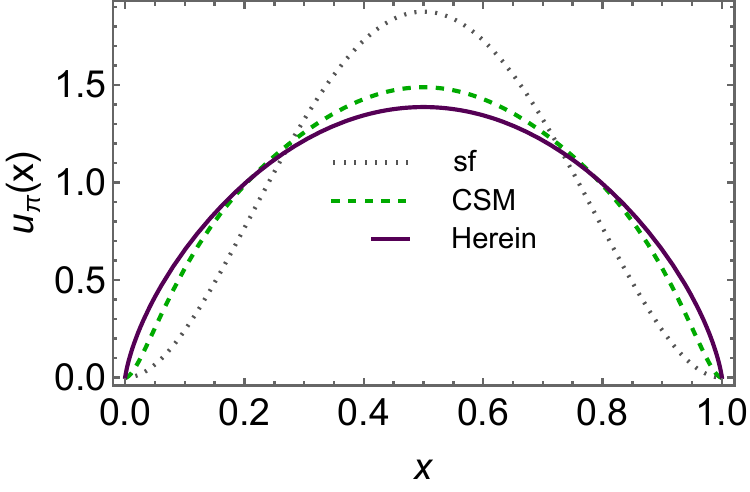}
\caption{\label{Fig3}
Pion valence quark DF.
Legend:
solid purple curve -- result obtained herein using ${\mathpzc a}(x)$ from Fig.\,\ref{Fig1} and Eq.\,\eqref{DFw};
dashed green curve -- CSM prediction \cite{Ding:2019lwe, Cui:2020tdf};
dotted grey curve -- scale free DF (see text).
}
\end{figure}


With the information now in hand, one readily arrives at the hadron scale DF depicted in Fig.\,\ref{Fig3}.  The endpoint behaviour of this $x\leftrightarrow (1-x)$ symmetric function is:
\begin{equation}
{\mathpzc u}_\pi(x) \stackrel{x\simeq 1}{=} 3.163 (1-x)\ln 1/(1-x) + \ldots\ .
\end{equation}

Once again, the image compares the symmetry-based result with the CSM prediction \cite{Ding:2019lwe, Cui:2020tdf}.
In this case, too, there is qualitative and semi-quantitative agreement; and, once more,
the differences owe to the omission of scaling violations in the monopole approximation to the pion elastic electromagnetic form factor, which are also expressed in a different endpoint behaviour:
${\mathpzc u}_\pi^{\rm CSM}(x) \approx 375.3 (1-x)^2$ on $x\simeq 1$.
Owing to its connection with QCD, the CSM prediction expresses the power law behaviour predicted by strong interaction theory
\cite{Brodsky:1994kg, Yuan:2003fs, Holt:2010vj, Cui:2021mom, Lu:2023yna}.
The other curve in Fig.\,\ref{Fig3} is the scale-free DF: ${\mathpzc q}_{\rm sf} = 30 x^2 (1-x)^2$, which is a DF analogue of $\varphi_{\rm asy}$.

Evidently, the symmetry properties discussed in Sec.\,2 ensure that, for any meson whose elastic electromagnetic form is well approximated by a monopole, the hadron scale valence quark DF is a dilated function.
Again, this is a model-independent result.

\begin{figure}[t]
\vspace*{-1ex}

\includegraphics[width=0.93\columnwidth]{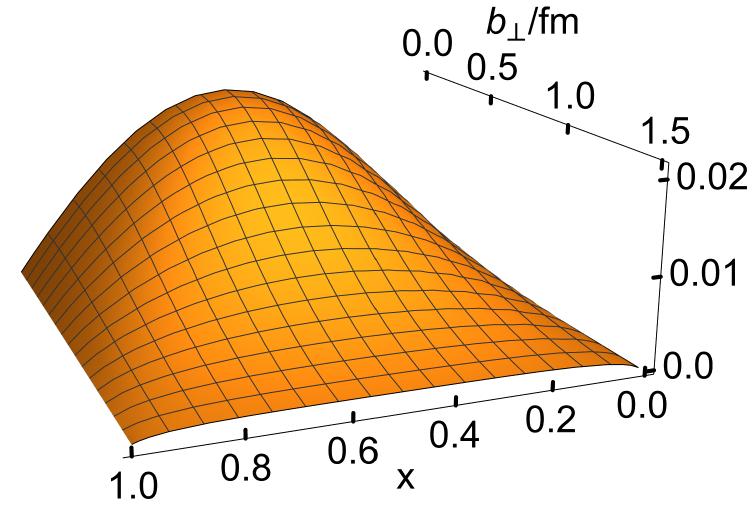}
\caption{\label{Fig4}
Hadron scale valence quark symmetrised-GPD in impact parameter space, Eqs.\,\eqref{EqGPD}, \eqref{HSGPDb}.
}
\end{figure}

Before providing a comparison with experiment, it is worth displaying the $\zeta_{\cal H}$ valence-quark symmetrised-GPD in impact-parameter space, Eqs.\,\eqref{EqGPD}, \eqref{HSGPDb}; see Fig.\,\ref{Fig4}.
Naturally, the result is symmetric under $x\leftrightarrow (1-x)$.
Moreover, owing to the absence of scaling violations in the monopole approximation to the electromagnetic form factor, then, at fixed-$x$, the $b_\perp$ profile is Gaussian with magnitude and range characterised by $m_M$.

\smallskip

\noindent\textit{6.\ Validation by experiment}\,---\,%
As noted in the Introduction, data relating to the pion valence quark DF are available \cite[E615]{Conway:1989fs, Wijesooriya:2005ir}.
Their interpretation remains controversial, with different phenomenological analysis methods yielding conflicting results \cite{Cui:2021mom}.
The existing data are from $\pi + W$ Drell-Yan reactions performed more than thirty-five years ago, at an average centre-of-mass energy $\zeta_5=5.2\,$GeV.
Relevant new data are expected to be obtained by the AMBER Collaboration \cite{Quintans:2022utc, Seitz:2023rqm} and, potentially, via experiments at Jefferson Laboratory \cite{Horn:2016rip} or an electron ion collider \cite{Chen:2020ijn, Arrington:2021biu}.
Meanwhile, comparisons can only be made with E615 data.

\begin{figure}[t]
\includegraphics[width=0.93\columnwidth]{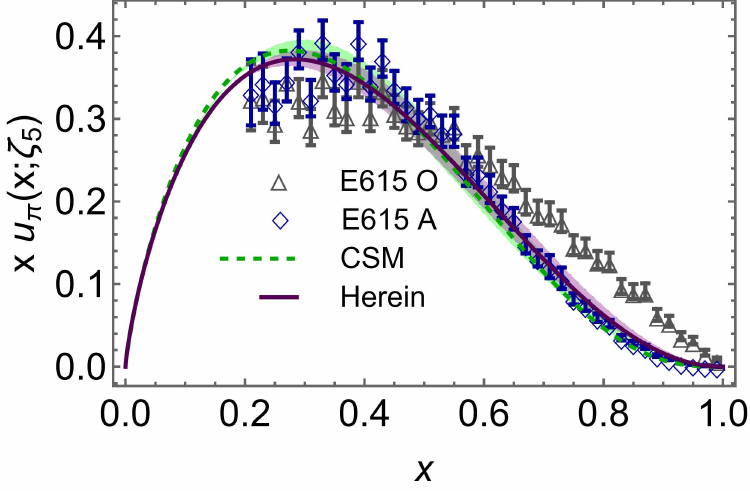}
\caption{\label{Fig5}
Pion valence quark DF at the scale, $\zeta_5 = 5.2\,$GeV.
Legend:
solid purple curve and band -- symmetry-driven result obtained herein, Eq.\,\eqref{DFw};
dashed green curve and band -- CSM prediction \cite{Ding:2019lwe, Cui:2020tdf};
open blue diamonds -- E615 data analysed as described in Refs.\,\cite{Aicher:2010cb, Chang:2014lva}, with soft-gluon resummation;
open grey up-triangles -- data inferred via leading-order analysis of the Drell-Yan cross section \cite{Conway:1989fs}
}
\end{figure}


In connecting $\zeta_{\cal H}\to \zeta_5$, we use the AO scheme \cite{Yin:2023dbw} for evolving the results in Fig.\,\ref{Fig3}.
To implement this, one only requires knowledge of the light-front momentum fraction carried by valence quarks at $\zeta_5$, since the hadron scale value is one-half by definition; see Eq.\,\eqref{DFsym}.
To be explicit, working with the DF Mellin moments:
\begin{equation}
\langle x^n \rangle_{{\mathpzc u}_{\pi}}^{\zeta}
= \int_{0}^{1} dx\, x^n {\mathpzc u}_{\pi}(x;\zeta)\,,
\end{equation}
then in the AO scheme:
\begin{equation}
\langle x^n \rangle_{{\mathpzc u}_{\pi}}^{\zeta_5} =
\langle x^n \rangle_{{\mathpzc u}_{\pi}}^{\zeta_{\cal H}}
[\langle 2 x \rangle_{{\mathpzc u}_{\pi}}^{\zeta_5}]^{\tilde\gamma_m^n},
\end{equation}
where the anomalous dimension $\tilde\gamma_m^n = -(3/8)(3+2/(n+1)/(n+2)-4 \sum_{k=1}^{n+1}1/k)$.
%
Namely, given the pion valence-quark hadron-scale DF, the value of every moment of this DF at $\zeta_5 > \zeta_{\cal H}$ is determined once the lowest nontrivial moment -- the momentum fraction -- is known at the new scale; and the pointwise behaviour of this DF is readily reconstructed from the set of all Mellin moments.

A unifying analysis of results obtained using continuum and lattice Schwinger function methods yields $\langle x \rangle_{{\mathpzc u}_\pi}^{\zeta_5} = 0.21 (1)$ \cite{Lu:2023yna}.
Working with this information, the AO procedure delivers the solid purple curve in Fig.\,\ref{Fig5}.
The associated bracketing band expresses the $\approx 5$\% uncertainty in the momentum fraction.
On $x\gtrsim 0.85$, the symmetry-based results are well described by a simple power-law: ${\mathpzc c}(1-x)^\beta$,
$\beta = 1.99\pm 0.06$, ${\mathpzc c}= 1.88 \mp 0.09$.
Similarly, the dashed green curve and band represent AO evolution of the CSM prediction: on $x\gtrsim 0.85$, they can be described by ${\mathpzc c}(1-x)^\beta$,
$\beta = 2.79\pm 0.08$, ${\mathpzc c}= 7.54 \mp 0.07$.
Evidently, since the differences are small, it is unlikely that any foreseeable experiment could achieve sufficient precision to enable analysis thereof to distinguish between the symmetry-based result obtained herein and the realistic CSM prediction; hence, for all intents and purposes, they are equivalent.
Importantly, both predictions unambiguously favour the analysis of E615 data described in Refs.\,\cite{Aicher:2010cb, Chang:2014lva}, which includes soft gluon resummation and, like the symmetry-based and CSM predictions, yields large-$x$ behaviour that is practically consistent with QCD expectations \cite{Brodsky:1994kg, Yuan:2003fs, Holt:2010vj, Cui:2021mom, Lu:2023yna}.

\smallskip

\noindent\textit{7.\ Conclusion}\,---\,%
Unveling the structure of Nature's most fundamental Nambu-Goldstone boson -- the pion -- is a high-priority goal of modern hadroparticle physics.
Using the theory of effective charges in QCD and the fact that the $Q^2$-dependence of the charged-pion elastic electromagnetic form factor is well approximated by a monopole, we delivered a model-independent, parameter-free prediction for the pion's valence-quark momentum-fraction probability distribution function [Fig.\,\ref{Fig5}].
%
%
Crucially, its behaviour on the valence domain of light-front momentum fraction, $x\gtrsim 0.2$, is practically indistinguishable from that of an array of predictions derived from QCD.
Our result confirms those obtained using continuum and lattice Schwinger function methods \cite{Cui:2021mom, Lu:2023yna}; hence may serve as additional motivation for future experiments that aim to resolve existing pion DF controversies \cite{Quintans:2022utc, Seitz:2023rqm, Horn:2016rip, Chen:2020ijn, Arrington:2021biu, Roberts:2021nhw}.

\smallskip

\noindent\textit{Acknowledgments}\,---\,%
%
Work supported by:
National Natural Science Foundation of China, grant no.\ 12135007;
and Spanish Ministry of Science and Innovation (MICINN), grant no.\ PID2022-140440NB-C22).

\smallskip

\noindent\textit{Appendix}\,---\,%
In order to expose the impact of the empirical monopole assumption, Eq.\,\eqref{MonoAssume}, we note that
\begin{subequations}
\label{A1}
\begin{align}
\frac{1}{(1+t)^{1+\epsilon}} & =
\int_{0}^{1}dy\, g(y) y^{t/2}, \label{A1a}\\
g(y) & = \frac{1}{2^{1+\epsilon}} \frac{1}{\surd y} [\ln 1/y]^\epsilon / \Gamma(1+\epsilon)\,,
\label{A1b}
\end{align}
\end{subequations}
where $t=Q^2/m_M^2$.  Now write $y=w_\epsilon(x)$, where this function has the properties stipulated in Eq.\,\eqref{wcons}, wherewith Eq.\,\eqref{A1a} maps into:
\begin{subequations}
\begin{align}
\frac{1}{(1+t)^{1+\epsilon}} &= \int_0^1 dx\, {\mathpzc u}_{\pi;\epsilon}(x) [w_\epsilon(x)]^{t/2}\,,\\
{\mathpzc u}_{\pi;\epsilon}(x) & = w_\epsilon^\prime(x) g(w_\epsilon(x))\,,
\end{align}
\end{subequations}
from which one immediately reads the associated GPD following the path to Eq.\,\eqref{Hfinal}.

At this point, the form of $w_\epsilon(x)$ is completely determined by Eq.\,\eqref{EqMaster} following the replacement
\begin{subequations}
\begin{align}
  [a(x)]^{h(x)} & \to \int_{0}^{[w_{\epsilon}(x)]^{h(x)}} dw_{\epsilon} g(w_\epsilon) \\
  & = \Gamma(1+\epsilon , -\tfrac{1}{2} {h(x)} \ln w_\epsilon(x) ) / \Gamma(1+\epsilon)\,,
\end{align}
\end{subequations}
where $\Gamma(a,b)$ is the (upper) incomplete gamma function.

Proceeding further, using the explicit form for $g(w_\epsilon(x))$, Eq.\,\eqref{A1b}, one has
\begin{subequations}
\begin{align}
{\mathpzc u}_{\pi;\epsilon}&(x)  = \tfrac{1}{2} \frac{w_\epsilon^\prime(x)}{\surd w_{\epsilon}(x)}
\tfrac{1}{2^\epsilon} [\ln 1/w_{\epsilon}(x)]^\epsilon / \Gamma(1+\epsilon) \\
& \stackrel{\mbox{Eq.\,\eqref{dew}}}{= } {\mathpzc u}_{\pi;0}(x) \tfrac{1}{2^\epsilon} [\ln 1/w_{\epsilon}(x)]^\epsilon / \Gamma(1+\epsilon)\,.
\end{align}
\end{subequations}

Consider now $\epsilon \simeq 0$, then Eqs.\,\eqref{wcons}, \eqref{aendpt} entail that
$w_\epsilon(x) \approx 1 - 2 a_1 (1-x)^2 \ln 1/(1-x)+ {\rm O}[(1-x)^4/\ln^2 \{1/(1-x)\}]$ on $x\simeq 1$; so,
\begin{equation}
{\mathpzc u}_{\pi;\epsilon}(x) =
{\mathpzc u}_{\pi;0}(x) a_1^{\epsilon} (1-x)^{2\epsilon} \ln^\epsilon [1/(1-x)]  / \Gamma(1+\epsilon)\,.
\end{equation}
Now, it is clear that an increase in the $Q^2$-power of form factor damping leads to a softer DF on $x\simeq 1$.  The discussion proceeds similarly on $x\simeq 0$.

This establishes the desired result; namely,
those minor differences which exist between the DA and DF results herein and CSM predictions thereof -- Figs.\,\ref{Fig2}, \ref{Fig3}, \ref{Fig5} -- are simply an expression of the absence of scaling violations in the monopole approximation to $F_\pi(Q^2)$, which can readily be corrected.
Consequently, regarding the principal features of our discussion, they are immaterial.



\providecommand{\newblock}{}

\end{document}